\documentclass[journal=jctcce,manuscript=article, layout=twocolumn]{achemso}

\usepackage{amsmath,amssymb}
\usepackage{graphicx}
\usepackage{color}
\usepackage[utf8]{inputenc}
\usepackage[colorlinks]{hyperref}
\usepackage{bm}

\usepackage[linesnumbered, ruled]{algorithm2e}
\usepackage[version=4]{mhchem}
\usepackage{physics}
\usepackage{qcircuit}
\setkeys{acs}{abbreviations=true}

\newcommand{\mr}[1]{\mathrm{#1}}
\newcommand{\mc}[1]{\mathcal{#1}}

\newcommand{\cnot}{\texttt{CNOT} }

\newcommand{\bmth}{\bm{\theta} }

\title{ADAPT-QSCI: Adaptive Construction of an Input State for Quantum-Selected Configuration Interaction}

\newcommand{\qunasys}{QunaSys Inc., Aqua Hakusan Building 9F, 1-13-7 Hakusan, Bunkyo, Tokyo 113-0001, Japan}
\newcommand{\handai}{Graduate School of Engineering Science, Osaka University, 1-3 Machikaneyama, Toyonaka, Osaka 560-8531, Japan}
\newcommand{\qiqb}{Center for Quantum Information and Quantum Biology, Osaka University, 1-2 Machikaneyama, Toyonaka, Osaka 560-0043, Japan}

\author{Yuya O. Nakagawa}
\email{nakagawa@qunasys.com}

\author{Masahiko Kamoshita}
\affiliation{\qunasys}

\author{Wataru Mizukami}
\affiliation{\qiqb}
\alsoaffiliation{\handai}

\author{Shotaro Sudo}
\author{Yu-ya Ohnishi}
\affiliation{Materials Informatics Initiative, RD technology and digital transformation center,
JSR Corporation, 3-103-9, Tonomachi, Kawasaki-ku, Kawasaki, Kanagawa, 210-0821, Japan}

\begin{document}

\begin{abstract}
We present a quantum-classical hybrid algorithm for calculating the ground state and its energy of the quantum many-body Hamiltonian by proposing an adaptive construction of a quantum state for the quantum-selected configuration interaction (QSCI) method.
QSCI allows us to select important electronic configurations in the system to perform CI calculation (subspace diagonalization of the Hamiltonian) by sampling measurement for a proper input quantum state on a quantum computer, but how we prepare a desirable input state has remained a challenge.
We propose an adaptive construction of the input state for QSCI in which we run QSCI repeatedly to grow the input state iteratively.
We numerically illustrate that our method, dubbed \textit{ADAPT-QSCI}, can yield accurate ground-state energies for small molecules, including a noisy situation for eight qubits where error rates of two-qubit gates and the measurement are both as large as 1\%.
ADAPT-QSCI serves as a promising method to take advantage of current noisy quantum devices and pushes forward its application to quantum chemistry.
\end{abstract}

\maketitle

\section{Introduction \label{sec:intro}}

With the rapid progress of quantum information technology in recent years, there appear various quantum devices that can host quantum computation without error correction.
Such devices are called noisy intermediate-scale quantum (NISQ) devices~\cite{Preskill2018}.
Quantum computers, including NISQ devices, are expected to be utilized in various fields of science and industry.
One of the most promising applications is the simulation of quantum many-body systems, such as quantum chemistry and condensed matter physics.

Variational quantum eigensolver (VQE)~\cite{Peruzzo2014,Tilly2022} is the most celebrated algorithm for exploiting NISQ devices to simulate quantum systems.
VQE is designed to calculate an approximate ground-state energy of a given Hamiltonian
by minimizing an energy expectation value of a trial quantum state that is realized on a quantum computer.
However, there are several practical difficulties in running VQE on the current NISQ devices.
For example, the number of gates required to generate a quantum state in VQE to solve large problems is often too large~\cite{OBrien2022purification}; it is desirable to be able to generate a state with as few gates as possible because of the noise involved in the gate execution of NISQ devices.
Another difficulty is that the number of measurements required to estimate the energy with sufficient accuracy is enormous.
Gonthier \textit{et al.}~\cite{Gonthier2022} claimed that as many as $10^{10}$ measurements (quantum circuit runs) would be required to accurately estimate the combustion energy of hydrocarbons.

With such problems with VQE becoming apparent, some of the authors in this study previously proposed a method not based on VQE to simulate quantum systems~\cite{kanno2023quantum}.
In this method, called quantum-selected configuration interaction (QSCI), a simple sampling measurement is performed on an input state prepared by a quantum computer.
The measurement result is used to pick up the electron configurations (identified with computational basis states) for performing the selected configuration interaction (CI) calculation on classical computers, i.e., Hamiltonian diagonalization in the selected subspace.
QSCI relies on a quantum computer only for generating the electron configurations via sampling, and the subsequent calculations to output the ground-state energy are executed solely by classical computers. 
This property leads to various advantages over VQE, for example, the number of measurements for energy estimation is small, the effect of noise in quantum devices can be made small, and so on.
It should be noted that QSCI still has a chance for exhibiting quantum advantage although the ground-state energy is calculated by classical computers because the sampling from a quantum state is classically hard for certain quantum states~\cite{arute2019quantum}.

The remaining challenge to take advantage of QSCI in various applications is that there is still no simple method to determine the input states of QSCI, which must contain important configurations for describing a ground state of a given Hamiltonian.
In this study, we propose a method to construct the input state of QSCI in an adaptive manner by iterative use of QSCI, making it possible to compute the ground-state energy of quantum many-body systems with fewer quantum gates and fewer measurement shots.
Specifically, similar to adaptive derivative-assembled pseudo-Trotter ansatz variational quantum eigensolver (ADAPT-VQE)~\cite{Grimsley2019,Tang2021} and qubit coupled cluster (QCC)~\cite{QCC2018,iQCC2020} based on VQE, we define a pool of operators that are generators of rotation gates for the input quantum state of QSCI, and select the best operator from the pool to lower the energy output by QSCI.
The selection of operators and the determination of rotation angles are performed solely by classical computation using classical vectors obtained by QSCI.
We repeatedly perform QSCI (with classical and quantum computers) and improve the input state of QSCI by adding a new rotation gate (determined with only classical computers).
The quantum computer only needs to repeatedly perform simple sampling measurements in the QSCI algorithm,
so the required quantum computational resources like the total number of measurements are expected to be greatly reduced compared with VQE-based methods.
As a numerical demonstration of our method, which we call ADAPT-QSCI, we perform quantum circuit simulations for calculating the ground-state energies of quantum chemistry Hamiltonians of small molecules.

This paper is organized as follows.
In Sec.~\ref{sec:preliminaries}, we review two methods on which our proposal is based: QSCI and ADAPT-VQE.
We explain our main proposal, ADAPT-QSCI, in Sec.~\ref{sec:proposal}.
Section~\ref{sec:numerics} is dedicated to the numerical demonstration of our proposed method.
We summarize our study and discuss possible extensions of our method and future work in Sec.~\ref{sec:conclusion}.
Details of the numerical calculation are explained in Supporting Information.

\section{Preliminaries \label{sec:preliminaries}}
In this section, we review two previous methods on which our proposal is based.
First, we explain quantum-selected configuration interaction (QSCI)~\cite{kanno2023quantum}.
Second, we review the ADAPT-VQE~\cite{Grimsley2019,Tang2021} algorithm, which is a proposal for iterative construction of quantum states for VQE.

\subsection{Review of quantum selected configuration interaction (QSCI)}
QSCI is a method to calculate eigenenergies and eigenstates of a given Hamiltonian $H$ by configuration interaction (CI) calculation whose configurational space is selected by quantum computers.
That is, one can calculate the eigenenergies and eigenstates by diagonalizing the Hamiltonian projected onto the subspace designated by the result of measurements on an input quantum state on a quantum computer.
Here we review the algorithm of QSCI for the case of finding the ground state and its energy.

Let us consider an $n$-qubit fermionic Hamiltonian $H_f$ and the corresponding qubit Hamiltonian $H$.
We assume the mapping between the qubit and fermion where each computational basis $\ket{i} \: (i=0,1,\cdots,2^n-1)$ in the qubit representation corresponds to one electronic configuration (Slater determinant).
For example, Jordan-Wigner~\cite{Jordan1928}, Parity~\cite{Bravyi2002, Seeley2012}, and Bravyi-Kitaev~\cite{Bravyi2002} transformation satisfy this property.

QSCI can be seen as one of the selected CI methods in quantum chemistry~\cite{helgaker2014molecular, bender1969pr,whitten1969jcp,huron1973jcp,buenker1974tca,buenker1975tca,nakatsuji1983cluster,cimiraglia1987jcc,harrison1991jcp,greer1995jcp,greer1998jcpss, evangelista2014jcp,holmes2016jctc,schriber2016jcp,holmes2016jctc2,tubman2016deterministic,ohtsuka2017jcp,schriber2017jctc,sharma2017semistochastic,chakraborty2018ijqc,coe2018jctc,coe2019jctc,abraham202jctc,tubman2020modern,zhang2020jctc,zhang2021jctc,chilkuri2021jcc,chilkuri2021jctc,goings2021jctc,pineda2021jctc,jeong2021jctc,seth2023jctc}.
In selected CI, the subspace is defined by choosing some set of configurations, $\mc{C} = \mr{span}\{ \ket{c_1}, \cdots, \ket{c_R} \}$, where $c_k \in \{0,1,\cdots,2^n-1\}$ are selected configurations and $R$ is the dimension of the subspace.
Then, the $R\times R$ projected Hamiltonian in the subspace whose components are $H^{\mc{C}}_{kl} = \mel{c_k}{H}{c_l}$ is constructed and diagonalized to yield the approximate ground state and its energy by classical computers.
Note that each matrix component $H^{\mc{C}}_{kl}$ can be efficiently calculated by classical computers thanks to the Slater-Condon rules.
How to choose the subspace that results in a good approximation of the exact eigenenergies and eigenstates is crucial to selected CI. 

QSCI exploits an ``input" quantum state $\ket{\Phi}$ and quantum computers to construct the subspace for selected CI.
We perform the projective measurement on the computational basis for the state $\ket{\Phi}$, called \textit{sampling}.
When $\ket{\Phi}$ is expanded in the computational basis as
\begin{equation} \label{eq:basis expansion}
\ket{\Phi} = \sum_{i=0}^{2^n-1} \alpha_i \ket{i},
\end{equation}
the projective measurement produces integers $i$ (or $n$-bit bitstrings) with the probability $|\alpha_i|^2$.
The number of repetitions of the projective measurement is called \textit{shot}.
After $N_s$ shots measurement, one obtains $N_s$ integers as $i_1, i_2, \cdots, i_{N_s} \in \{0,1,\cdots,2^n-1\}$.

The algorithm of QSCI is outlined as follows:
\begin{enumerate}
 \item Prepare the input state $\ket{\Phi}$ on a quantum computer and repeat the projective measurement on the computational basis $N_s$ times.
 \item Compute the occurrence frequency of each integer (configuration) $i$ in the results of $N_s$ shots measurement: $f_i = n_i/N_s$, where $n_i$ is the number of $i$ appearing in the measurement result $i_1, \cdots, i_{N_s}$.
 \item Choose the $R$ most-frequent configurations, $r_1, \cdots, r_R \in \{0,1,\cdots,2^n-1\}$, and define the subspace $\mc{S} = \mr{span}\{\ket{r_1}, \cdots, \ket{r_R} \}$.
 \item Perform selected CI calculation, or diagonalization of the projected Hamiltonian in the subspace $\mc{S}$, by classical computers. This gives the approximate ground state and ground-state energy of the Hamiltonian.
\end{enumerate}

The crucial point for the success of QSCI is the choice of the input state $\ket{\Phi}$, which must contain important configurations to describe the exact ground state with large weight $|\alpha_i|^2$.
Ideally, the exact ground state $\ket{\psi_\mr{GS}}$ of the Hamiltonian $H$ itself is a candidate for such input state because the weights $|\alpha_i|^2$ for important configurations in $\ket{\psi_\mr{GS}}$ are, almost by definition, large.
One can pick up important configurations by the projective measurement for $\ket{\psi_\mr{GS}}$.
Therefore, it is reasonable to use quantum states with lower energy expectation value $\ev{H}{\Phi}$ as input states of QSCI, which must resemble the exact ground state.
Following this consideration, the original proposal of QSCI~\cite{kanno2023quantum} utilized a quantum state generated by VQE with a loose optimize condition in some numerical and hardware experiments.
In this study, we propose a method to construct a proper input state for QSCI by iteratively growing the input state by repeating the QSCI calculation in Sec.~\ref{sec:proposal}.

We finally comment on the choice of $R$, a dimension of the subspace.
As we perform the diagonalization of $R \times R$ projected Hamiltonian in the subspace by classical computers, $R$ must be smaller than what is capable of classical computers.
The original QSCI paper~\cite{kanno2023quantum} studied this point and claimed that $R$ required to calculate energies of \ce{Cr2} described by 50 qubits with the accuracy of $10^{-3}$Ha is about $10^3$.
This suggests that there are some molecules in which selected CI (and thus QSCI) is helpful to study their energy even though the full-space diagonalization for $O(2^{n})$-dimensional matrix is impossible. 
In this study, we assume that $R$ is not so large that we can diagonalize the projected Hamiltonian by classical computers within a reasonable amount of time.

\subsection{Review of ADAPT-VQE}
In VQE~\cite{Peruzzo2014, Tilly2022}, a quantum state called \textit{ansatz} state is defined by a parameterized quantum circuit,
\begin{equation}
 \ket{\psi(\bmth)} = U(\bmth)\ket{\psi_0},
\end{equation}
where $\bmth=(\theta_1, \cdots, \theta_M)$ are parameters and $\ket{\psi_0}$ is some initial state.
VQE aims at minimizing the expectation value $\ev{H}{\psi(\bmth)}$ with respect to the parameters $\bmth$.
One can obtain the approximate ground state and energy at the optimized parameters $\bmth^*$ as $\ket{\psi(\bmth^*)}$ and $\ev{H}{\psi(\bmth^*)}$, respectively.

The parameterized circuit $U(\bmth)$ is one of the most important factors for the success of VQE because it determines the expressibility of the ansatz and hence the quality of the resulting approximate ground state and energy~\cite{Sim2019}.
Complicated parameterized circuits containing a large number of quantum gates have high expressibility, but are difficult to execute on NISQ devices due to noisy gate operations.
The compact (or shallow) parameterized circuit with a high capability of expressing ground states of Hamiltonians of interest is demanded.

ADAPT-VQE~\cite{Grimsley2019} is a method to construct such a shallow circuit $U(\bmth)$ for the ansatz of VQE.
First, a ``operator pool" $\mathbb{P}=\{P_1,P_2,\cdots P_T\}$ is defined.
The operators $P_j$ must be Hermitian and we consider the rotational gate generated by it, $e^{i\theta P_j}$, for the construction of the ansatz.
The algorithm of ADAPT-VQE is as follows.
\begin{enumerate}
 \item Define an operator pool $\mathbb{P}=\{P_1, \cdots, P_T\}$ and an initial state $\ket{\psi_0}$. Set $k=0$.
 \item Evaluate the expectation value 
\begin{equation} \label{eq:VQE grad}
 g_j = \ev{i[H, P_j]}{\psi_k}   
\end{equation}
 for all $P_j \in \mathbb{P}$ by using quantum computers, where $[H, P_j]=HP_j -P_j H$ is a commutator.
 \item Choose the operator in the pool that has the largest $|g_j|$ and we denote it $P_{t_k}$. If $|g_{t_k}|$ is smaller than some threshold, the algorithm stops.
 \item Grow the ansatz for VQE as $\ket{\psi_{k+1}(\theta_0, \cdots, \theta_k)} = e^{i\theta_k P_{t_k}} \cdots e^{i\theta_0 P_{t_0}} \ket{\psi_0} $. Perform VQE for this ansatz, i.e., minimize the energy expectation value
\begin{align*}
 \ev{H}{\psi_{k+1}(\theta_0, \cdots, \theta_k)}            
\end{align*}
with respect to $\theta_0, \cdots, \theta_k$, using quantum computers.
The optimized state and energy are denoted $\ket{\psi_{k+1}}$ and $E_k^{\mr{(VQE)}}$, respectively.
\item  Set $k \gets k+1$ and go back to Step 2. 
\end{enumerate}
The important point of the ADAPT-VQE algorithm is Step 2, where the operators in the pool are ranked by $g_j$, which is in fact identical to the gradient of the energy expectation value for a trial state $e^{i\theta P_j}\ket{\psi_k}$,
\begin{equation}
 g_j = \left. \frac{d}{d\theta} \mel{\psi_k}{e^{-i\theta P_j}He^{i\theta P_j}}{\psi_k} \right|_{\theta=0}.
\end{equation}
Choosing the most effective operator in the pool judged by this gradient and growing the ansatz for VQE iteratively can make the resulting ansatz shallow and accurate.

The operator pool is chosen as the single and double excitations of electrons in the original proposal of ADAPT-VQE~\cite{Grimsley2019}.
Later a simpler pool made of single Pauli operators was proposed in a method called ``qubit ADAPT-VQE"~\cite{Tang2021}, and it can create the shallow ansatz with high accuracy for quantum chemistry Hamiltonians of small molecules.
It should also be noted that a method called qubit coupled-cluster (QCC)~\cite{QCC2018} and its iterative version (iterative QCC)~\cite{iQCC2020} employed the similar ranking of the operators in the pool and proposed the pool made of single Pauli operators.

\section{ADAPT-QSCI \label{sec:proposal}}
Here, we explain our proposal, adaptive construction of the input state of QSCI (ADAPT-QSCI), for finding the ground state and its energy of quantum many-body Hamiltonian.
Our proposal is similar to ADAPT-VQE, but we repeatedly use QSCI instead of VQE with several modifications of the algorithm.
The algorithm of ADAPT-QSCI is described as follows (details of these Steps are explained later).

\begin{enumerate}
  \item Define an operator pool $\mathbb{P}=\{P_1, \cdots, P_T\}$ composed of single Pauli operators and an initial state $\ket{\Phi_0}$.
  Set the number of iterations $k=0$.
 \item Perform QSCI with the input state $\ket{\Phi_k}$ with $N_s$ shots and the maximum dimension of the subspace $R$.
 QSCI generates the $R_k (\leq R)$-dimensional subspace $\mc{S}_k = \mr{span} \left\{ \ket{r_1^{(k)}}, \cdots, \ket{r_{R_k}^{(k)}} \right\}$, the projected Hamiltonian $H^{\mc{S}_k}$ in the subspace, its smallest eigenvalue $E_k$, and the associated $R_k$-dimensional classical eigenvector $\bm{c}_k$.
 If the energy $E_k$ is converged compared with the energies in some previous iterations, the algorithm stops.
 \item Evaluate the value of
 \begin{equation} \label{eq:qsci grad}
  h_j = \ev{i[H,P_j]}{\bm{c}_k}
 \end{equation}
 for all operators $P_j$ in the pool $\mathbb{P}$ by \textit{classical computers}.
 Here, $\ket{\bm{c}_k}$ is a state corresponding to the classical vector $\bm{c}_k$,
 \begin{equation}
  \ket{\bm{c}_k} = \sum_{l=1}^{R_k} (\bm{c}_k)_l \ket{r_l^{(k)}},
 \end{equation}
 where $(\bm{c}_k)_l$ is the $l$-th component of $\bm{c}_k$.
 The value of $h_j$ can be calculated by projecting the Hermitian operator $i[H, P_j]$ onto the subspace $\mc{S}_k$ and computing the expectation value of it for (classical vector) $\bm{c}_k$, both of which are efficiently accomplished by classical computers.
 Similar to ADAPT-VQE, $h_j$ is identical to the gradient of energy expectation value for a state $e^{i\theta P_j} \ket{\bm{c}_k}$,
 \begin{equation}
    h_j = \left. \frac{d}{d\theta} \mel{\bm{c}_k}{e^{-i\theta P_j}He^{i\theta P_j}}{\bm{c}_k} \right|_{\theta=0}.
 \end{equation}
\item Choose the operator in the pool that has the largest $|h_j|$ and we denote it $P_{t_k}$. 
 \item Determine a parameter for a new input state for QSCI.
 We consider a (hypothetical) state $e^{i\theta_k P_{t_k}} \ket{\bm{c}_k}$ and minimize the energy expectation value
\begin{align} \label{eq:QSCI cost}
 f_k(\theta_k) = \mel{\bm{c}_k}{e^{-i\theta_k P_{t_k}} H e^{i\theta_k P_{t_k}}}{\bm{c}_k}
\end{align}
with respect to $\theta_k$.
This can also be performed only by classical computers as we have
\begin{equation}
\begin{aligned}
 e^{-i\theta_k P_{t_k}} H e^{i\theta_k P_{t_k}} = \cos^2{\theta_k} H + \sin^2{\theta_k} P_{t_k}HP_{t_k} \\
 + i\sin{\theta_k}\cos{\theta_k} [H, P_{t_k}]
\end{aligned}    
\end{equation}
because $P_{t_k}$ is a Pauli operator satisfying $P_{t_k}^2 = I$.
It is possible to evaluate $f_k(\theta_k)$ for any value of $\theta_k$ by projecting $H, P_{t_k}HP_{t_k}$, and $i[H, P_{t_k}]$ onto the subspace $\mc{S}_k$ and computing the expectation values of them for the state $\ket{\bm{c}_k}$, which can be efficiently carried out by classical computers.
Since $f_k(\theta_k)$ is a simple trigonometric function, we can compute the exact minimum of $f_k(\theta_k)$ from these expectation values~\cite{Nakanishi2019}.
Let $\theta_k^*$ be the optimal value of $\theta_k$.
\item  Define a new state for QSCI as $\ket{\Phi_{k+1}} = e^{i\theta_k^* P_{t_k}} \ket{\Phi_k}$.
Set $k \gets k+1$ and go back to Step 2. 
\end{enumerate}

The crucial part of the algorithm is from Steps 3 to 5, where we try to minimize the energy expectation value of the state $e^{i\theta P_j}\ket{\bm{c}_k}$ for $P_j$ in the pool by ranking the operators by their gradient $h_j$ and optimizing the angle $\theta$.
We expect that this procedure leads also to lowering the energy expectation value of $e^{i\theta_k P_j }\ket{\Phi_k}$, the input state for QSCI at the next iteration.
The input state with a small energy expectation value is favorable for QSCI as we explained in Sec.~\ref{sec:preliminaries}.
This expectation is not rigorously supported by analytical arguments, but we see this is the case up to some extent in numerical simulation in the next section.
Note that the input state for QSCI at each iteration, $\ket{\Phi_k}$, is considered to be classically intractable although its counterpart, $\bm{c}_k$ or $\ket{\bm{c}_k}$, can be handled by classical computers because the former is represented as a $2^n$-dimensional vector whereas the latter is just a $R$-dimensional vector; therefore, using quantum computers at Step 2 in the ADAPT-QSCI algorithm is still meaningful to gain a possible quantum advantage.

ADAPT-QSCI uses quantum computers only in Step 2 of the algorithm for performing QSCI, which leads to several favorable properties for executing on quantum computers, especially NISQ devices.
First, since QSCI needs a simple projective measurement on the computational basis, the number of measurements required to perform the algorithm is much smaller than the conventional VQE~\cite{Gonthier2022, kanno2023quantum} and ADAPT-VQE.
Especially, the reduction of the number of measurements compared with ADAPT-VQE is expected when we rank the operators in the pool.
The ranking is performed by calculating $h_j$ in Eq.~\eqref{eq:qsci grad} using only classical computers in ADAPT-QSCI while the value of $g_j$ in Eq.~\eqref{eq:VQE grad} must be estimated by usual expectation value measurements on quantum computers in ADAPT-VQE.
Evaluating $g_j$ for all operators in the pool by quantum computers is one of the bottlenecks of ADAPT-VQE~\cite{majland2023fermionic}.
Second, ADAPT-QSCI is expected to be more noise-robust than conventional VQE-based methods because the energy is computed by classical computers and not directly estimated by the results of noisy quantum circuit measurements.
We numerically illustrate the noise-robustness of ADAPT-QSCI in the next section.
Third, the iterative nature of ADAPT-QSCI may allow us to construct a shallow input state for QSCI similarly as (qubit) ADAPT-VQE provides a shallow ansatz state for VQE.
Namely, as a result of ADAPT-QSCI, we obtain the input state for QSCI,
\begin{equation}
 \ket{\Phi_{m}} = e^{i\theta_{m-1}^* P_{t_{m-1}}} \cdots  e^{i\theta_0^* P_{t_0}} \ket{\Phi_0},
\end{equation}
where $m$ is the number of iterations when the algorithm stops.
Our method can solve the challenge of QSCI in that there is no practical and systematic construction of its input state. 

We have additional comments on the algorithm.
First, the parameters $\theta_0^*, \cdots, \theta_{k-1}^*$ are not changed when we determine $\theta_k^*$ at the iteration $k$ in our algorithm whereas ADAPT-VQE optimizes all parameters at each iteration.
Second, the choice of the operators in Steps 3 and 4 can be made by a different criterion.
For example, calculating minimal values of $\mel{\bm{c}_k}{e^{-i\theta P_j} H e^{i\theta P_j}}{\bm{c}_k}$ with respect to $\theta$ for all $P_j$ in the pool is feasible with classical computers so that one can choose the operator $P_j$ based on these minimal values (see Ref.~\cite{feniou2023adaptive} for the same strategy in ADAPT-VQE).
Third, we stress that the classical computation required in our algorithm is the matrix-vector multiplication for at most $R$ dimension.
The state $e^{i\theta_k P_{t_k}}\ket{\bm{c}_k}$ considered from Steps 3 to 5 possibly has larger-than-$R$ non-zero components, but we avoid a direct treatment of it by an algebraic calculation of $e^{-i\theta_k P_{t_k}} H e^{i\theta_k P_{t_k}}$.
Finally, Majland \textit{et al.}~\cite{majland2023fermionic} recently proposed using the same projective measurement on the computational basis as ours for ranking the operators in the pool of ADAPT-VQE.
Their proposal differs from ours in that they considered performing the usual VQE to optimize and output the energy of the Hamiltonian.

\section{Numerical demonstration \label{sec:numerics}}
In this section, we present a numerical demonstration of our algorithm for Hamiltonians of small systems in quantum chemistry.
In noiseless quantum circuit simulation, we first compare the number of \cnot gates to create the quantum state and the number of measurement shots of our method with those of ADAPT-VQE.
We note that we still consider the statistical fluctuation of the results of measurements.
We then show that our method works well in molecules with electronic correlations by taking nitrogen molecule~\ce{N2} as an example.
In noisy quantum circuit simulation for an eight-qubit system where two-qubit gates and measurement have errors as large as 1\%, we exemplify that ADAPT-QSCI still gives accurate energy.
This illustrates the noise-robustness of our method to some extent.

\subsection{Setup}
We consider the following small systems: hydrogen chains \ce{H4} and \ce{H6} with bond length 1\AA{} and nitrogen molecule \ce{N2} with various bond lengths (1.1\AA, 1.3\AA, 1.5\AA, 1.7\AA, and 2.0\AA).
We use the STO-3G basis set and construct molecular Hamiltonians of electrons by using the Hartree-Fock orbitals computed by PySCF~\cite{Sun2018,Sun2020}.
Jordan-Wigner transformation~\cite{Jordan1928} is used to map the electronic Hamiltonians into qubit one by using OpenFermion~\cite{McClean2020}.
The quantum circuit simulation is performed by Qulacs~\cite{Suzuki2021} interfaced by QURI-Parts~\cite{quri-parts}.

The operator pool for numerical calculation is chosen by following the original proposal of qubit ADAPT-VQE~\cite{Tang2021}, as we described in Sec.~S1 of Supporting Information.
The number of $R$ for ADAPT-QSCI is systematically chosen by referring to the exact ground-state wavefunction so that QSCI being input the exact ground state can yield sufficiently accurate energy (see Sec.~S2 of Supporting Information).
We made this choice because we want to evaluate the performance of ADAPT-QSCI without taking the effect of the value of $R$ into account.
When performing QSCI in the ADAPT-QSCI algorithm, we drop configurations whose observed frequency is less than $10^{-10}$ as well as those who do not possess the correct number of electrons and the total $z$-component of spin of electrons; after such post selection, we take $R$ configurations at most to define the subspace of QSCI.
We also count the number of \cnot gates to create the state $\ket{\Phi_k}$ in ADAPT-QSCI or $\ket{\psi_k}$ in ADAPT-VQE under the standard decomposition of the Pauli rotation gate $e^{i\theta P}$ by assuming all-to-all connectivity among the qubits, as details are presented in Sec.~S3 of Supporting Information.
Since \cnot gate is one of the largest sources of error and costly operations in current NISQ devices, the number of \cnot gates indicates how easy or hard to create a quantum state on NISQ devices.
The convergence of ADAPT-QSCI is detected when the difference between QSCI energies $E_k$ and $E_{k-1}$ gets smaller than $10^{-5}$ Hartree throughout this section.

\subsection{Noiseless simulation}

\begin{figure*}
 \includegraphics[width=0.49\linewidth]{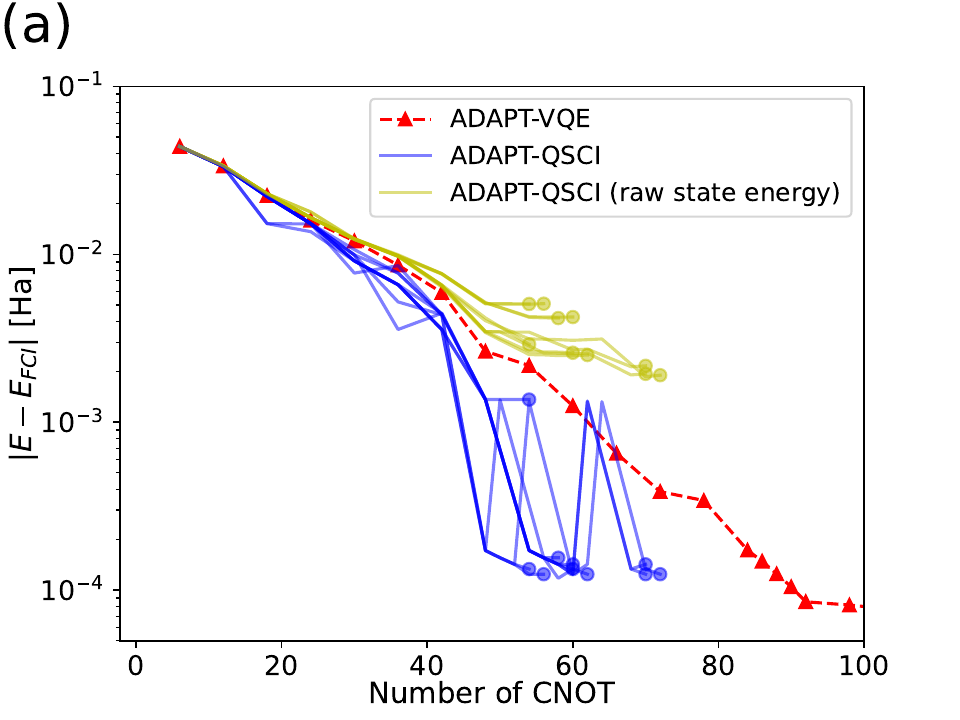}
 \includegraphics[width=0.49\linewidth]{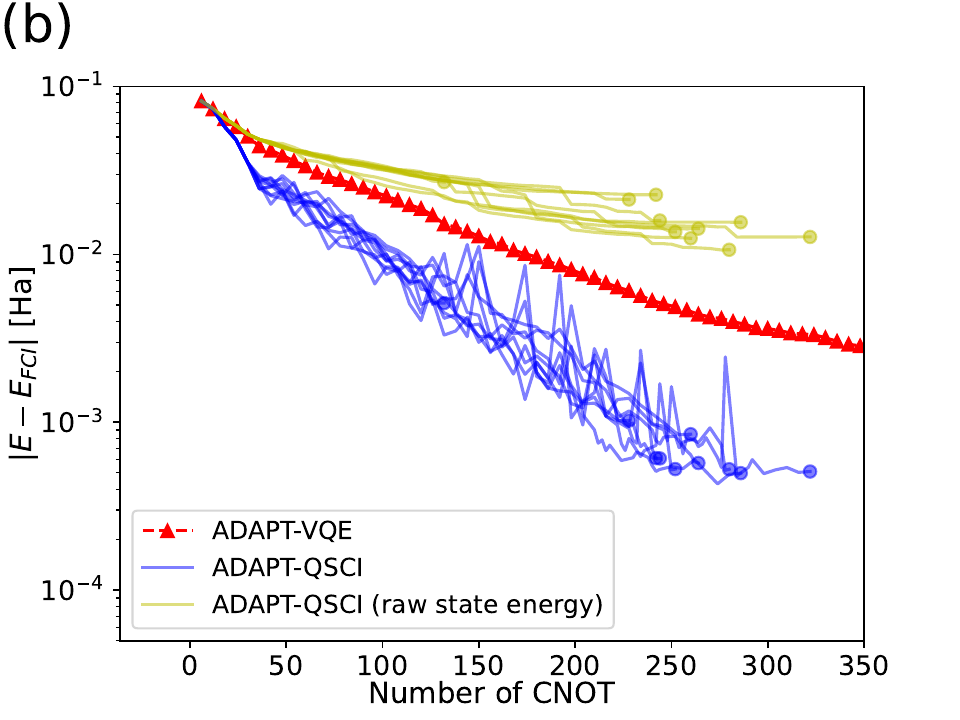}
 \caption{Numerical result of ADAPT-QSCI and qubit ADAPT-VQE with the same operator pool (Eq. (S1) in Supporting Information) for hydrogen chains \ce{H4} (a) and \ce{H6} (b).
 Blue lines represent the QSCI energy $E_k$ in ADAPT-QSCI at each iteration and blue dots do the converged energy for ten runs of the algorithm.
 Yellow lines represent the energy expectation value of the input state of QSCI $\ev{H}{\Phi_k}$ and yellow dots do the values at the iteration when $E_k$ converges.
 Red dots and lines represent the VQE energy $E_k^\mr{(VQE)}$ in qubit ADAPT-VQE at each iteration.
 \label{fig:comparison with VQE}}
\end{figure*}

In the noiseless simulation of ADAPT-QSCI, we simulate the quantum state $\ket{\Phi_k}$ at each step of the algorithm without considering any error in circuit execution and measurement.
We still consider the statistical fluctuation of the projective measurement due to the finite number of shots when performing QSCI.
Specifically, we simulate $\ket{\Phi_k}$ without error and generate the result of the projective measurement (integers, or electron configurations) by random numbers obeying the distribution defined by the state through the expansion like Eq.~\eqref{eq:basis expansion}.

\subsubsection{Comparison with ADAPT-VQE}
First, we compare our method with qubit ADAPT-VQE~\cite{Tang2021}, focusing on the number of gates for generating quantum states and the number of measurement shots required for running the algorithm.
Simulation of ADAPT-VQE is performed without any noise and the statistical fluctuation due to the finite number of shots; we use the exact expectation value like $\ev{i[H, P_j]}{\psi_k}$ or $\ev{H}{\psi_k}$ in ADAPT-VQE.
The same operator pool is used for both algorithms.
The optimization of VQE in ADAPT-VQE is performed by the Broyden–Fletcher–Goldfarb–Shanno (\texttt{BFGS}) algorithm implemented in SciPy~\cite{2020SciPy-NMeth}.
Hydrogen chains \ce{H4} and \ce{H6} with atomic bond length 1.0\AA, respectively corresponding to 8 and 12 qubits, are considered.
The number of configurations $R$ (and shots $N_s$) for ADAPT-QSCI is $R=14$ ($N_s=10^5)$ for \ce{H4} and $R=119$ ($N_s=5\times 10^6)$ for \ce{H6}.
We ran 10 trials for ADAPT-QSCI to see the effect of the randomness of the measurement results in QSCI.

The result is shown in Fig.~\ref{fig:comparison with VQE}, where we plot the number of \cnot gates to create $\ket{\Phi_k}$ of ADAPT-QSCI and $\ket{\psi_k}$ of ADAPT-VQE at each iteration versus the energy difference to the exact ground-state energy.
For both \ce{H4} and \ce{H6}, ADAPT-QSCI gives accurate energy with less CNOT gates.
It is seen that the energy expectation value of the state in ADAPT-QSCI, $\ev{H}{\Phi_k}$, gets also smaller as the iteration of ADAPT-QSCI proceeds, indicating that the input state of QSCI is improved along the iterations.

We also estimate the total number of measurement shots to run both algorithms.
The average number of shots used in ADAPT-QSCI for ten runs in Fig.~\ref{fig:comparison with VQE} is 
\begin{equation} \label{eq:ADAPT-QSCI shot ave.}
\begin{aligned}
N_\mr{tot}^\mr{(ADAPT-QSCI)} = \: & 1.26\times 10^6 \:\: (\text{\ce{H4}}), \\
& 2.25 \times 10^8 \:\:(\text{\ce{H6}}),    
\end{aligned}
\end{equation}
respectively.
For ADAPT-VQE, we roughly estimate the total number of shots to run the whole algorithm based on that required to estimate the energy expectation value $\ev{H}{\psi_\mr{GS}}$ for the exact ground state $\ket{\psi_\mr{GS}}$ \textit{once},
denoted $N_{E,\mr{once}}^\mr{(VQE)}$.
As we explain in Sec.~S4 of Supporting Information,
we need
\begin{equation} \label{eq: VQE E once}
\begin{aligned}
N_{E,\mr{once}}^\mr{(VQE)} = \: &  1.02 \times 10^6 \:\: (\text{\ce{H4}}), \\
&  3.96 \times 10^6 \:\:(\text{\ce{H6}}),
\end{aligned}
\end{equation}
to make the standard deviation of the estimate of $\ev{H}{\psi_\mr{GS}}$ smaller than $10^{-3}$ Hartree.
On the other hand, the number of iterations $k$ of ADAPT-VQE to reach the VQE energy whose precision is $|E_k^{\mr{VQE}} - E_\mr{exact}| \leq 10^{-3}$ Hartree is $k=10$ for \ce{H4} and  $k>150$ for \ce{H6} in our numerical simulation, so one has to evaluate the energy expectation value at least $k+1=11$ (\ce{H4}) and $k+1>151$ (\ce{H6}) times even if we ignore a lot of evaluations of the energy expectation values required to perform VQE optimization at each iteration.
From these considerations, we can estimate a rough and possibly loose estimate for the total number of shots to run ADAPT-VQE as
\begin{equation} \label{eq:VQE shot est.}
\begin{aligned}
N_\mr{tot}^\mr{(VQE)} \gtrsim \: & 11 N_{E,\mr{once}}^\mr{(VQE)}= 1.1 \times 10^7  \:\: (\text{\ce{H4}}), \\
& 151 N_{E,\mr{once}}^\mr{(VQE)} = 6.0 \times 10^8   \:\:(\text{\ce{H6}}).
\end{aligned}
\end{equation}
In addition to this, the evaluation of $g_j$ (Eq.~\eqref{eq:VQE grad}) on a quantum computer is needed for all operators in the pool whose number is 164 (\ce{H4}) and 1050 (\ce{H6}) in our setup.
Therefore, the actual number of shots must be larger than Eq.~\eqref{eq:VQE shot est.}, and this clearly illustrates the efficiency of measurement shots in ADAPT-QSCI.

\subsubsection{\ce{N2} with various bond length}
\begin{figure}
 \includegraphics[width=0.8\linewidth]{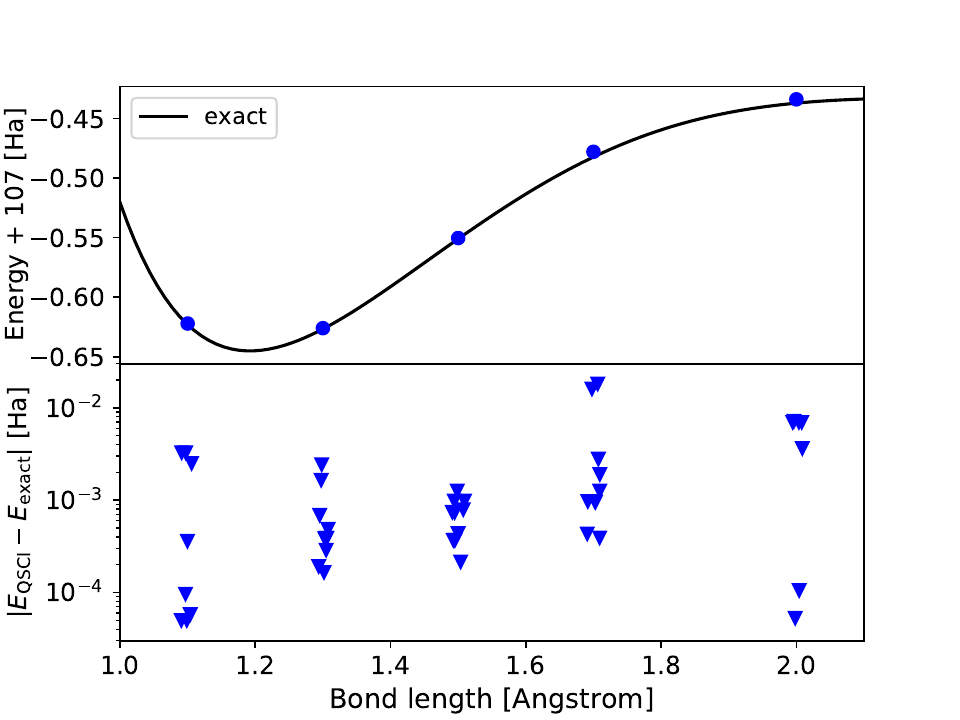}
 \caption{The result of ADAPT-QSCI along with the dissociation curve of \ce{N2}.
 We run ADAPT-QSCI ten times for the Hamiltonians of bond lengths 1.1\AA, 1.3\AA, 1.5\AA, 1.7\AA, and 2.0\AA.
 The blue dots in the upper panel represent the mean energy output of ten runs. 
 The data points in the lower panel represent the difference to the exact energy and are jittered in the horizontal axis so that all data points are visible.
 \label{fig:N2 result}
 }
\end{figure}

Next, we evaluate the performance of ADAPT-QSCI in systems with multiple-bond dissociation. 
We consider \ce{N2} molecule with five bond lengths: 1.1\AA, 1.3\AA, 1.5\AA, 1.7\AA, and 2.0\AA, along the dissociation curve of \ce{N2}. 
This is one of the typical benchmark problems in quantum chemistry.
We employ the active space of six orbitals and six electrons closest to HOMO and LUMO, so the Hamiltonian is described by 12 qubits.
The number of configurations $R$ for ADAPT-QSCI is set to $R=52$, obtained by applying the procedure in Sec.~S2 of Supporting to the system of 2.0\AA.
The number of shots $N_s$ is taken as $N_s = 5 \times 10^6$.
We run ADAPT-QSCI ten times for each bond length.
We note that we do not present the result of ADAPT-VQE for this case because it requires a much larger number of shots than ADAPT-QSCI, as shown above for \ce{H4} and \ce{H6} molecules.

The result is shown in Fig.~\ref{fig:N2 result}.
For all bond lengths examined, ADAPT-QSCI gives accurate energies whose differences to the exact energy are around between $10^{-2}$ Hartree and $10^{-4}$ Hartree.
This exemplifies the ability of ADAPT-QSCI to tackle systems with electron correlations.
We note that the results of long bond lengths (1.7\AA{} and 2.0\AA) are slightly worse than the others, and we attribute it to the fact that the exact ground states at long bond lengths may require more configurations to describe them (see details in Sec.~S5 of Supporting Information).
Several options, such as the use of other orbitals than the Hartree-Fock, can be considered to mitigate this.

\subsection{Noisy simulation}
\begin{figure}
 \includegraphics[width=0.8\linewidth]{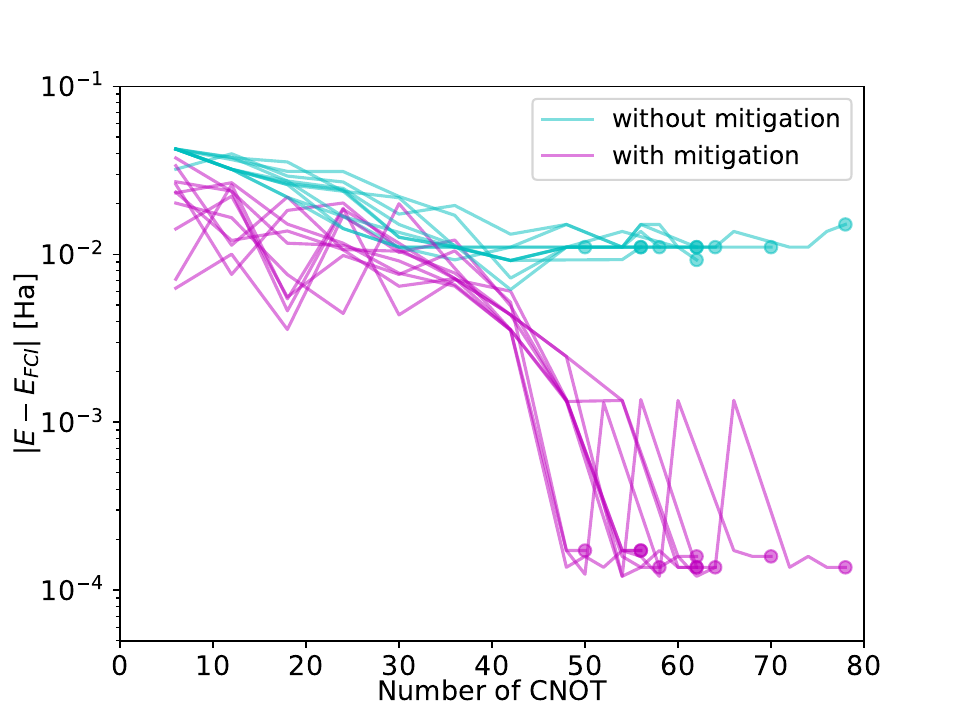}
 \caption{Noisy simulation of ADAPT-QSCI for \ce{H4} molecule.
 The magenta lines represent the QSCI energy $E_k$ with error-mitigated measurement results at each iteration for ten trials,
 and the cyan lines do the QSCI energy $E_k$ with unmitigated measurement results.
 The error rates for two-qubit gate ($p_{2q}$) and measurement ($p_m$) are set to $p_{2q}=p_m=1\%$.
 \label{fig:noisy sim.}
 }
\end{figure}

The currently available quantum devices have errors in their operations such as gate applications and measurements.
Here, we perform the simulation of ADAPT-QSCI in noisy situations to check the validity of the method on such noisy devices. 

As sources of noise in the simulation, we consider gate error for two-qubit gates and error for the measurement.
Our noise model is characterized by the error rate for \cnot gates $p_{2q}$ and the error rate for the measurement $p_m$.
Specifically, we apply the single-qubit depolarizing channel to all involving qubits after the application of Pauli rotation gate $e^{i\theta P}$ in ADAPT-QSCI, where $P$ is either a two-qubit or four-qubit Pauli operators in the operator pool.
The strength of the depolarizing channel is determined by reflecting the decomposition of $e^{i\theta P}$ into \cnot gates and the error rate for \cnot gates $p_{2q}$.
The measurement error is implemented by inserting $X$ gate (bit-flip) with the probability $p_m$ for all qubits just before the projective measurement on the computational basis. 
Further details are explained in Sec.~S6 of Supporting Information.

In noisy simulation, we utilize error mitigation techniques~\cite{Temme2017, Endo2018practical}, specifically, the digital zero-noise extrapolation~\cite{digitalZNE} and the measurement error mitigation~\cite{Qiskit2023, Maciejewski2020}, to alleviate the effect of the noise.
Both methods aim at recovering the noiseless result of quantum measurement from the noisy result by consuming additional computational costs (the number of measurements).
The error mitigation techniques have typically been developed for expectation values of observables, but here we apply them to the result of the projective measurement on the computational basis.
The observed (noisy) frequency of observing a bit $i=0,1,\cdots,2^n-1$ in QSCI, denoted $f_i^\mr{(obs)}$, is extrapolated to the noiseless one $\tilde{f}_i$ in our error mitigation.  
See Sec.~S6 of Supporting Information for concrete formulations.

The noisy simulation result for \ce{H4}, described by eight qubits, is presented in Fig.~\ref{fig:noisy sim.}.
The number of configurations and measurements for QSCI at each iteration of ADAPT-QSCI is set to $R=14$ and $N_s=10^5$, respectively.
We observe that ADAPT-QSCI with error mitigation provides accurate energy whose difference to the exact one is smaller than $10^{-3}$ Hartree for all ten trials, similarly as the noiseless simulation in Fig.~\ref{fig:comparison with VQE}(a).
The average number of measurements for one run of the algorithm is $4.06\times 10^6$, reflecting the overhead of the error mitigation compared with the noiseless case~\eqref{eq:ADAPT-QSCI shot ave.}.
We also plot the QSCI energy using the unmitigated result of the projective measurement, $f_i^\mr{(obs)}$, in Fig.~\ref{fig:noisy sim.}.
Our result of the noisy simulation in the 8-qubit system demonstrates the noise-robustness of ADAPT-QSCI even when the gate noise and the measurement noise are both 1\%, comparable to the current noise level of NISQ devices.
We note that the noisy simulation and real device experiment of QSCI were already performed in Ref.~\cite{kanno2023quantum}, but the input state of QSCI was determined by classical simulation of VQE for the target system that is not scalable due to exponentially large classical computational cost.
Our illustration in this section determines the input state of QSCI based on the (simulated) result of quantum devices in a scalable manner (as far as $R$ is not exponentially large) so that it is considered more practical.

\section{Summary and outlook \label{sec:conclusion}}

In this study, we propose a quantum-classical hybrid algorithm to compute the ground state of quantum many-body Hamiltonian and its energy by adaptively improving the initial state of QSCI.
Our method, named ADAPT-QSCI, allows us to calculate ground-state energies using quantum states with a small number of gates and requiring a small number of measurement shots.
Numerical simulations for the Hamiltonians in quantum chemistry show that the ADAPT-QSCI can efficiently and accurately calculate the ground-state energy with a small number of two-qubit gates and a small number of measurements.
Moreover, in simulations with noise, which is inevitable in NISQ devices, accurate energies were calculated even when errors in the quantum gates and measurement results were as large as 1\%.
Our method is not only promising as a realistic usage of NISQ devices, but also as a simple preparation method for input states in quantum phase estimation, which enhances the potential applications of quantum computers.

Interesting directions for future work of this study include the extension to excited states like the original QSCI proposal~\cite{kanno2023quantum}.
A large-scale experiment using actual NISQ devices will also be important to further verify the effectiveness of our method.
Finally, we note that a state constructed by ADAPT-QSCI can be leveraged as an initial state of quantum phase estimation since the state will have larger overlaps with the exact ground state than the Hartree-Fock state itself.

\section*{Acknowledgements}
 YON thanks Keita Kanno for helpful comments on the manuscript. 
 WM is supported by funding from the MEXT Quantum Leap Flagship Program (MEXTQLEAP) through Grant No. JPMXS0120319794, and the JST COI-NEXT Program through Grant No. JPMJPF2014. The completion of this research was partially facilitated by the JSPS Grants-in-Aid for Scientific Research (KAKENHI), specifically Grant Nos. JP23H03819 and JP21K18933.

\bibliography{ref} 

\end{document}


In this Supporting Information, we explain the details of numerical calculation in Sec.~4 of the main text.

\section{Operator pool \label{subsec:operator pool}}
The operator pool used in Sec.~4 of the main text is defined as follows by referring to qubit ADAPT-VQE~\cite{Tang2021}.
When the system is described by spin-independent $N_\mr{MO}$ molecular orbitals, there are $n=2N_\mr{MO}$ fermion operators $a_{0,\uparrow}, a_{0,\downarrow}, \cdots, a_{N_\mr{MO}-1,\uparrow}, a_{N_\mr{MO}-1,\downarrow}$ and their conjugates, where $a_{i,\sigma}$ corresponds to the annihilation operator of electron in $i$-th molecular orbital with spin $\sigma$.
Through Jordan-Wigner transformation, we label the indices of qubits so that $(0,\uparrow),(0,\downarrow), \cdots, (N_\mr{MO}-1,\uparrow), (N_\mr{MO}-1,\downarrow) $ correspond to $0,1,\cdots,n-1$.
In this ``up-down-up-down-..." notation, the operator pool $\mathbb{P}$ is defined by
\begin{equation}
\label{eq:pool}
\begin{aligned}
 \mathbb{P}= &\{X_{2i}Y_{2j}, X_{2i+1}Y_{2j+1} \mid 0\leq i < j \leq n/2-1\} \: \cup \\
 & \{ X_p X_q X_r Y_s, X_p X_q Y_r X_s, X_p Y_q X_r X_s, Y_p X_q X_r X_s  \\ 
 & \mid 0 \leq p < q < r < s \leq n-1, \\ & p+q+r+s \equiv 0 \text{ mod 2}\}.
\end{aligned}
\end{equation}
$XY$ terms in the pool correspond to generalized single electron excitations $c_{i,\sigma}^\dag c_{j,\sigma} - c_{j,\sigma}^\dag c_{i,\sigma}$ with ignoring $Z$ operators (Jordan-Wigner strings) between $2i(+1)$ and $2j(+1)$.
We also neglect YX terms that have the same indices as XY terms since they are related by the swap $X \leftrightarrow Y$~\cite{Tang2021}.
Similarly, $XXXY, XXYX, XYXX$, and $YXXX$ terms are inspired by generalized double electron excitations, and the condition $p+q+r+s\equiv 0 \text{ mod 2}$ mimics the conservation of $z$-component of the spin $S_z$ in the double excitations.

\section{Number of configurations used in ADAPT-QSCI \label{subsec:determine R}}
The number of configurations $R$ used in ADAPT-QSCI in Sec.~4 of the main text is determined similarly as Ref.~\cite{Kohda2022}.
We compute the exact ground state $\ket{\psi_\mr{GS}}$ of the Hamiltonian and expand it in the computational basis as $\ket{\psi_\mr{GS}} = \sum_i \alpha_{\mr{GS},i} \ket{i}$.
We sort the indices $i=0,1,\cdots,2^n-1$ in descending order of the weight $|\alpha_{\mr{GS},i}|^2$ and denote the sorted indices $s_0, s_1,\cdots,s_{2^n-1}$ (i.e., $|\alpha_{\mr{GS},s_0}|^2 \geq \cdots \geq |\alpha_{\mr{GS},s_{2^n-1}}|^2$).
We then define $R_\delta$ with $0<\delta\leq 1$ as the smallest $R$ that satisfies 
\begin{equation}
 \sum_{j=0}^{R-1} |\alpha_{\mr{GS},s_j}|^2 \geq 1 - \delta.
\end{equation}
In other words, the sum of $R_\delta$-largest amplitudes of the exact ground-state wavefunction is larger than $1 - \delta$.
The value of $\delta$ can be viewed as the infidelity of the (unnormalized) approximation of the exact ground-state wavefunction with retaining $R_\delta$-largest amplitudes.

We set $\delta=10^{-4}$ to determine the value of $R$ in Sec.~4 of the main text.
We numerically observe that QSCI with this choice of $R$ gives accurate energy whose difference to the exact energy is smaller than $10^{-3}$ Hartree when the input state of QSCI is the exact ground state.
Therefore, if the quantum state generated by ADAPT-QSCI is close to the exact ground state, the error of its energy is expected to be also small. 

\section{Counting the number of \cnot gates \label{subsec:cnot decomp}}
\begin{figure}
\includegraphics[width=\linewidth]{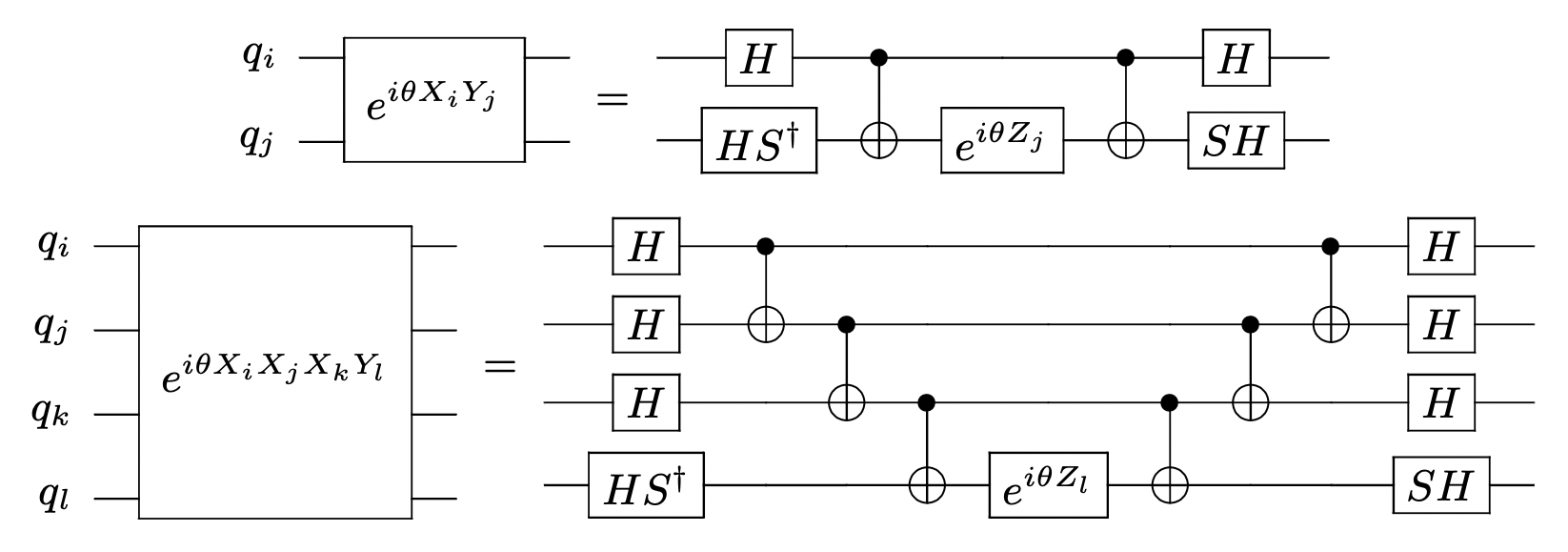}
\caption{Decomposition of multi-qubit Pauli rotations $e^{i\theta X_i Y_j}$ and $e^{i\theta X_i X_j X_k Y_k}$ into \cnot gates and single-qubit rotations. $q_i$ denotes $i$-th qubit.
\label{fig:gate-transpile}}
\end{figure}

We count the number of \cnot gates to create the state $\ket{\Phi_k}$ of ADAPT-QSCI and $\ket{\psi_k}$ of ADAPT-VQE as follows.
Both $\ket{\Phi_k}$ and $\ket{\psi_k}$ are composed of Pauli rotation gates $e^{i\theta P_2}$ or $e^{i\theta P_4}$, where $P_2 \, (P_4)$ is a two-qubit (four-qubit) Pauli operator in our numerical simulations, so we consider the number of \cnot gates to realize $e^{i\theta P_2}$ and $e^{i\theta P_4}$.
Note that the initial states $\ket{\Phi_0}$ and $\ket{\psi_0}$ in our numerical calculation are the Hartree-Fock states that are product states requiring no two-qubit gates.
As depicted in Fig.~\ref{fig:gate-transpile}, the rotation gate $e^{i\theta P_2}$ is decomposed into two \cnot gates and five single-qubit rotation gates.
Similarly, $e^{i\theta P_4}$ is decomposed into six \cnot gates and nine single-qubit rotation gates.

In the numerical simulations in Sec.~4 of the main text, we count the number of \cnot gates for creating $\ket{\Phi_k}$ and $\ket{\psi_k}$ by simply summing up all contributions from $e^{i\theta_k P_{t_k}}, \cdots, e^{i\theta_0 P_{t_0}}$.
This means that we assume the all-to-all connectivity among the qubits.
It should be also noted that there is room for optimizing the number of \cnot gates in our estimation, such as canceling some \cnot gates belonging to different rotation gates.

\section{Estimate of the number of shots for ADAPT-VQE \label{subsec:adaptvqe est.}}
When one estimates the energy expectation value $\ev{H}{\psi}$ for the Hamiltonian $H$ and a state $\ket{\psi}$ in VQE, the Hamiltonian is decomposed into the sum of Pauli operators:
\begin{equation} \label{eq: Ham.}
 H = \sum_{l=1}^L d_l P_l,
\end{equation}
where $d_l$ is a real coefficient, $P_l$ is the Pauli operator, and $L$ is the number of Pauli operators.
Furthermore, the Pauli operators are divided into groups whose components are mutually commutating to reduce the number of measurements for estimating the energy expectation value.
In Sec.~4 of the main text, we employ \texttt{SortedInsertion} method introduced in Ref.~\cite{Crawford2021efficientquantum} to evaluate the number of measurements to estimate the energy expectation value $\ev{H}{\psi_\mr{GS}}$, where $H$ is the Hamiltonian describing \ce{H4} or \ce{H6} and $\ket{\psi_\mr{GS}}$ is its exact ground state.
\texttt{SortedInsertion} method groups the Pauli operators in the Hamiltonian by sorting them by the amplitudes of their coefficients, and it was argued that this method exhibits a small number of measurements for various molecular Hamiltonian.  
We assume the optimal allocation of the measurement shots for the groups and estimate the number of measurements to realize the standard fluctuation of $10^{-3}$ Hartree for estimating the energy expectation value, using Eq.~(5) of Ref.~\cite{Crawford2021efficientquantum}.
The result is shown in Eq.~(12) of the main text.

\section{Distribution of amplitudes for \ce{N2} molecules \label{subsec:N2 distribution}}
\begin{figure}
    \includegraphics[width=\linewidth]{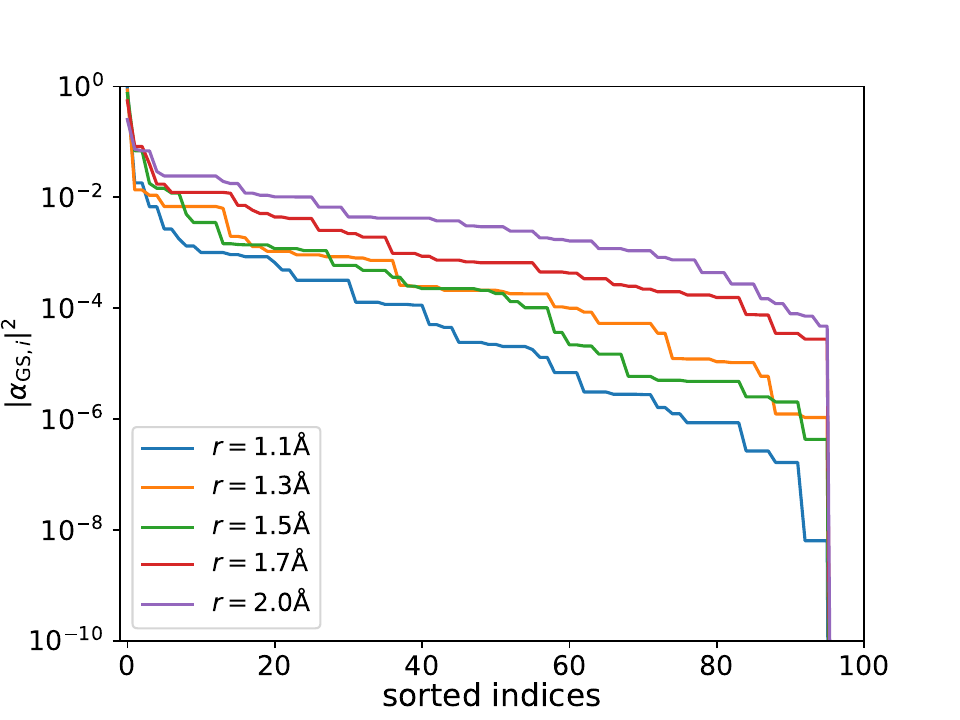}
    \caption{Distribution of the amplitudes of the exact ground-state wavefunction of the Hamiltonian for \ce{N2} molecule with various bond lengths.
    We sort the amplitudes in descending order and the horizontal axis represents the sorted indices for each bond length.
    \label{fig:N2 dist.}
    }
\end{figure}

We show the distribution of amplitudes $|\alpha_{\mr{GS},i}|^2$ for the exact ground state of the Hamiltonian for \ce{N2} molecule, $\ket{\psi_\mr{GS}}=\sum_i \alpha_{\mr{GS},i} \ket{i}$, in Fig.~\ref{fig:N2 dist.}.
The wavefunctions for long bond lengths have a broader distribution (slowing decay) of the amplitudes.
This property possibly results in the slightly worse performance of ADAPT-QSCI for long bond lengths because more configurations are needed to explain the exact ground state (and its energy) for such cases.

\section{Noise model and error mitigation \label{subsec:noise model}}
\begin{figure}
$$
\Qcircuit @C=0.8em @R=0.8em {
\lstick{q_i} & \multigate{1}{e^{i\theta P_2}} & \gate{D_i(p')} & \qw\\
\lstick{q_j} & \ghost{{e^{i\theta P_2}}} &\gate{D_j(p')} & \qw
}
\hspace{2.5em}
\Qcircuit @C=0.8em @R=0.8em {
\lstick{q_i} & \multigate{3}{e^{i\theta P_4}} & \gate{D_i(p'')} & \qw \\
\lstick{q_j} & \ghost{e^{i\theta P_4}} & \gate{D_j(p'')} & \qw \\
\lstick{q_k} & \ghost{e^{i\theta P_4}} & \gate{D_k(p'')} & \qw \\
\lstick{q_l} & \ghost{e^{i\theta P_4}} & \gate{D_l(p'')} & \qw
}
$$
\caption{Gate noise applied in the noisy simulation in Sec.~4 of the main text.
$D_q(p)$ is the single-qubit depolarizing noise on qubit $q$ with the probabilities $p', p''$ defined in Eq.~\eqref{eq:noise p definition}.
\label{fig:noise-def}}
\end{figure}
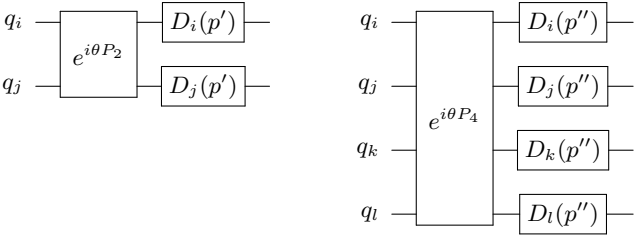

\subsection{Noise model}
We employ the following noise model for obtaining the result shown in Fig.~3 in Sec.~4 of the main text.
We consider the gate error induced by \cnot gates and the measurement error occurring at the projective measurement on the computational basis in QSCI.
We ignore the gate error of single-qubit gates for simplicity since their error is much smaller (typically one or two orders of magnitudes) than that of two-qubit gates in the current NISQ devices.
In our setup, a quantum circuit for creating the input state of QSCI state $\ket{\Phi_k}$ is composed of Pauli rotation gates $e^{i\theta P}$, where $P$ is either two-qubit Pauli operators or four-qubit Pauli operators in the operator pool.
Our noise model applies the single-qubit depolarizing channel $D_q(p)$ to all involving qubits $q$ after each $e^{i\theta P}$ is applied:
\begin{equation}
 D_q(p): \rho \mapsto (1-p) \rho + \frac{p}{3} \left( X_q \rho X_q + Y_q \rho Y_q + Z_q \rho Z_q \right),
\end{equation}
where $\rho$ is a state to which $e^{i\theta P}$ is applied, $X_q, Y_q$ and $Z_q$ are single-qubit Pauli operators on qubit $q$, and $p$ is the probability for the channel (see Fig.~\ref{fig:noise-def}).
The probability $p$ of the depolarizing channel is determined by referring to the decomposition of $e^{i\theta P}$ into \cnot gates and single-qubit gates.
As explained in Sec.~\ref{subsec:cnot decomp}, the rotation gate $e^{i\theta P}$ is decomposed into two \cnot gates when $P$ is two-qubit Pauli operators and six \cnot gates when $P$ is four-qubit Pauli operators.
We choose the probability $p'$ ($p''$) after the application of two-qubit (four-qubit) Pauli rotation as
\begin{align} \label{eq:noise p definition}
\begin{aligned}
 (1-p')^2 = (1-p_{2q})^2 \Leftrightarrow \: & p' = p_{2q}, \\
 (1-p'')^4 = (1-p_{2q})^6 \Leftrightarrow \: & p'' = 1 - (1 -p_{2q})^{3/2} \\ & \:\:\: (\approx 3p_{2q}/2),
\end{aligned}
\end{align}
where $p_{2q}$ is the error rate for \cnot gates.
These equations mean that the fidelity after applying two (six) \cnot gates becomes the same as that after applying the two (four) depolarizing channels.
Although this model does not perfectly describe the gate noise in actual quantum hardware, we expect that it reflects the essential feature of the noise and is enough to investigate the noise-robustness of ADAPT-QSCI.

In addition to the gate error noise above, our noise model introduces the measurement error by inserting $X$ gates (bit-flip noise) to all qubits before the projective measurement on the computational basis is performed.
The probability for inserting $X$ gates is independent of each other and set to $p_m$.

\subsection{Error mitigation}
We utilize two error mitigation techniques in the noisy simulation.
Let us consider performing the projective measurement on the computational basis for the input state $\ket{\Phi_k}$ at iteration $k$ of ADAPT-QSCI. 
The observed (noisy) frequency of observing a bit $i=0,1,\cdots,2^n-1$ is denoted $f_i^\mr{(obs)}$.
The goal of our error mitigation is to predict the noiseless result $\tilde{f}_i$ for the same state $\ket{\Phi_k}$ from noisy results and utilize it in QSCI calculation.

The first technique is a digital zero-noise extrapolation (ZNE)~\cite{Temme2017,Endo2018practical,digitalZNE}.
We execute the projective measurement on the computational basis for a quantum state $\ket{\Phi^{(3)}_k} = U_k U_k^\dag U_k \ket{0}$, where $U_k$ is defined by $\ket{\Phi_k}=U_k \ket{0}$, with $N_s$ shots.
We denote the frequency of measuring a bit $i$ for $\ket{\Phi_k^{(3)}}$ by $f_i^{(\mr{obs},3)}$.
Intuitively, the ``strength" of the noise in the measurement result $f_i^{(\mr{obs},3)}$ is three times larger than that in $f_i^{(\mr{obs})}$.
Therefore, the mitigated frequency $f_i^{(\mr{ZNE})}$ is calculated by the linear extrapolation: 
\begin{equation} \label{eq: ZNE formula}
 f_i^{(\mr{ZNE})} = \frac{3f_i^{(\mr{obs})} - f_i^{(\mr{obs},3)}}{2}. 
\end{equation}
The total number of measurements for each iteration of ADAPT-QSCI becomes $2N_s$ when we perform ZNE.

The second technique for error mitigation is read-out error mitigation (REM)~\cite{Qiskit2023, Maciejewski2020}.
Suppose that the correct probability of measuring $n$-digit integers $i=0,1,\cdots,2^n-1$ for the projective measurement on a given $n$-qubit quantum state is $p_i^\mr{(exact)}$ and the observed probability affected by the measurement noise is $p_i^\mr{(obs)}$.
We perform REM by assuming the relationship
\begin{equation}
 p_i^\mr{(obs)} = \sum_{j=0}^{2^n-1} A_{i,j} p_j^\mr{(exact)},
\end{equation}
where $A_{i,j}$ is a $2^n \times 2^n$ matrix called the calibration matrix.
We further assume that the calibration matrix $A$ is in the form of tensor product:
\begin{equation}
 A_{i_1 \cdots i_n, j_1 \cdots j_n} = A^{(1)}_{i_1,j_1} \cdot A^{(2)}_{i_2,j_2} \cdot \cdots \cdot A^{(n)}_{i_n,j_n},
\end{equation}
where $i_1 \cdots i_n$ and $j_1 \cdots j_n$ are binary representation of the integers $i$ and $j$, respectively, and $A^{(q)}$ is a $2\times 2$ calibration matrix for the qubit $q$.
Namely, $A^{(q)}$ is defined by 
\begin{equation}
 p_{i_q}^{(q,\mr{obs})} = \sum_{j_q=0,1} A^{(q)}_{i_q j_q} p_{j_q}^{(q,\mr{exact})},
\end{equation}
where $p_{i_q}^{(q,\mr{obs})}$ and $p_{i_q}^{(q,\mr{exact})}$ is the marginal probability of measuring $i_q$ for the qubit $q$ in noisy and noiseless situations, respectively. 
This equation means that the measurement error occurs at each qubit independently.
The exact probability is recovered by 
\begin{equation} \label{eq:REM formula}
\begin{aligned} 
 p_i^\mr{(exact)} &= \sum_{j=0}^{2^n-1} (A^{-1})_{i,j} p_j^\mr{(obs),}  \\
 (A^{-1})_{i,j} &= (A^{(1)})^{-1}_{i_1,j_1} \cdots  (A^{(n)})^{-1}_{i_n,j_n},
\end{aligned}
\end{equation}
in this scheme.
Note that this protocol is not scalable in that we need the inverse and multiplication of the $2^n \times 2^n$ matrix.
Nevertheless, we can use the scalable version of REM for large systems~\cite{Nation2021,yang2022}.

In the numerical simulation for Fig.~3 of the main text, we first estimate the calibration matrix $A^{(q)}_{i_q,0} (A^{(q)}_{i_q,1})$ for $q=1,\cdots,n$ by preparing the state $\ket{0}$ ($\ket{2^{q-1}}$) and measuring the probability of observing a bit $i_q=0,1$ for $q$-th qubit using $N_s$ shots, before starting the iterations of ADAPT-QSCI.
The total number of measurements to estimate the calibration matrix is $2nN_s$.
Then, at each iteration, the zero-noise-extrapolated frequency $f_i^{(\mr{ZNE})}$ [Eq.~\eqref{eq: ZNE formula}] is used as a observed frequency $p_i^\mr{(obs)}$ to calculate frequency $p_i^\mr{(exact)}$ in Eq.~\eqref{eq:REM formula}.
Finally, we perform the post-selection of an observed bit $i$ by setting $p_i^\mr{(exact)}=0$ for the bit $i$ whose corresponding electronic configuration does not have the correct number of electrons and the total $z$-component of the spin.
The post-selected frequency is utilized as the mitigated frequency $\tilde{f}_i$ in QSCI in the main text.

\bibliography{ref_SI} 